\documentclass[sigconf]{acmart}

\usepackage{subcaption}

\AtBeginDocument{%
  \providecommand\BibTeX{{%
    \normalfont B\kern-0.5em{\scshape i\kern-0.25em b}\kern-0.8em\TeX}}}

\setcopyright{acmcopyright}
\copyrightyear{2021}
\acmYear{2021}

\acmConference[ICOOOLPS '21]{ICOOLPS '21: Workshop on Implementation, Compilation, Optimization of OO Languages, Programs and Systems}{June 13, 2021}{Aarhus}

\begin{document}

\title{Userfault Objects: Transparent Programmable Memory}

\author{Konrad Siek}
\authornote{Both authors contributed equally to this research.}
\email{siekkonr@fit.cvut.cz}
\orcid{0000-0002-3599-2164}
\affiliation{%
  \institution{Faculty of Information Technology\\Czech Technical University}
  \streetaddress{Thákurova 9}
  \city{Prague}
  \country{Czech Republic}
  \postcode{160 00}
}

\author{Colette Kerr}
\email{kerrcole@fit.cvut.cz}
\affiliation{%
  \institution{Faculty of Information Technology\\Czech Technical University}
  \streetaddress{Thákurova 9}
  \city{Prague}
  \country{Czech Republic}
  \postcode{160 00}
}

\renewcommand{\shortauthors}{Siek and Kerr}



\maketitle

\section{Introduction}

Most objects are straightforward, but some object harbor secrets.  While most
objects are collections of assorted fields bundled with methods that operate on
them, occasionally an object is a transparent fa\c{c}ade providing an
abstraction over an underlying complex system.  Sometimes the fa\c{c}ade is
pierced and bad things happen. An example follows.

This is a definition of an object in the R language representing a sequence of
elements 1--10.

\begin{verbatim}
    simple <- as.integer(c(1,2,3,4,5,6,7,8,9,10))
\end{verbatim}

\noindent
Internally, this object is a simple vector with a header and a body consisting
of all of its member values.  The values can be accessed directly. For example,
the \verb|simple[i]| operator retrieves the the \verb|i|th indexed element by
accessing the memory at an offset from the end of the header.
However, the same sequence can be expressed as using the following simpler
syntax.

\begin{verbatim}
    magic <- 1:10
\end{verbatim}

\noindent
The \verb|magic| vector outwardly appears to be the same as the \verb|simple|
vector. However, internally the vector only contains two values---the beginning
and end of the range---and the values of the sequence are calculated on demand.
Then, \verb|magic[i]| is redefined to run the function calculating the value,
instead of accessing an offset. 

This alternative representation of vectors in R (ALTREP) \cite{Bec20} allows
implementing custom, even user-defined, back-ends to vectors while providing a
compatible API to ordinary R vectors. The advantage of this is the flexibility
of semantics and internal representation that allows implementing file-backed
persistent vectors, larger than memory vectors, and fast sequences with low
memory overheads. The disadvantage is that since the internal layout of ALTREP
vectors is so different from the layout of R vectors, the entire R runtime
needed retooling to handle them.
%
%

The abstraction ALTREP vectors present to the user can be pierced by
introspection.  While many languages provide mechanisms for observing the
internals of objects, in R this is perhaps easier than most.  R itself and many
R packages are written in C, so R provides a C API that allows packages to
interface with the runtime internals and runtime objects. This exposes the
layout of objects to external programmers, who are known to circumvent
prescribed API functions in favor of direct memory accesses into vectors.
ALTREP vectors defend themselves against this by materializing if the pointer
to the body of a vector is accessed via an API function. On the other hand, the
problem persists in general, as a sufficiently stubborn programmer may reach
into a vector via pointer arithmetic without reference to the API at all,
inadvertently dispelling the ALTREP abstraction and introducing segmentation
faults or subtle memory bugs.

There are a number of frameworks and runtime mechanisms providing similar
fa\c{c}ades without runtime support too. In R there are numerous libraries
providing transparent larger-than-memory vectors (matter \cite{BV17}, ff \cite{KEW13}, bigstatr \cite{PAZB18}, disk frame
\cite{df},  etc.) or abstractions
over SQL databases (dbplyr \cite{dbplyr}), in addition to ALTREP. Abstracting
frameworks are also found in other languages, e.g. Remote Objects \cite{RMI} in
Java and Dask data frames \cite{rock15} in Python. These can be introspected
into by the application of nefarious means.

We attempt to create completely transparent abstractions by exploring a
different approach We introduce a framework for Userfault Objects (UFOs).
\footnote{\url{https://github.com/PRL-PRG/UFOs}} 
\;%
UFOs expose an area of virtual memory to the program in some host language. This area is populated with the
representation of the object using the layout and contents that the host
language is expecting, but this is done lazily. Specifically, when an access to
the memory inside the object occurs, the UFO framework communicates with the
operating system (i.e.  with the Linux Kernel via \verb|userfaultfd|) to
materialize and populate a section of memory. The population procedure is
performed by a custom (user-defined) function which provides a specific slice
of the object. The population function can provide contents of the object by
calculating it or retrieving it from persistent storage (e.g. by parsing a CSV
file or running SQL queries), a remote site, or other external sources.
The ability to process data on the fly as it is being read, as well as to have
no backing persistent storage at all distinguishes UFOs from memory mapped
files.

\section{UFO core framework}

\begin{figure}[t]
  \centering
  \includegraphics[width=\linewidth]{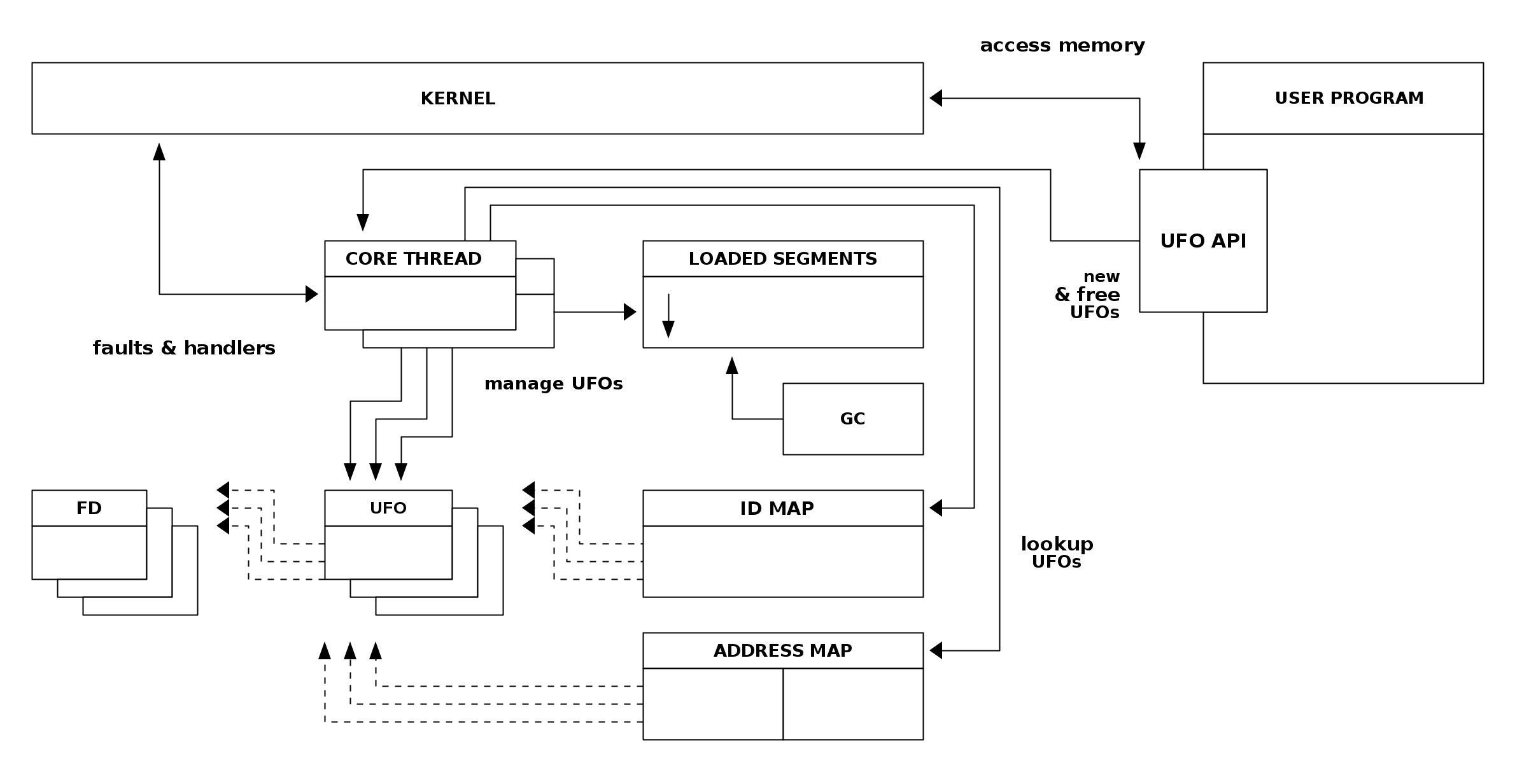}
  \caption{UFO core framework architecture.}
  \label{fig:ufo-core}
\end{figure}

Our proof-of-concept implementation consists of two layers: a language agnostic
core framework and a language specific API. This section describes the former.
\emph{UFO core} interacts directly with the operating system and manages the
creation and destruction of individual UFOs. It also handles reading and updating them.
The framework discharges its responsibilities via two cooperating subsystems: the
\emph{event API} and the \emph{page fault loop}, each running in a separate
operating system thread.
The event API is exposed as a façade through which UFOs can be created or freed.
The UFO API calls these functions directly.
The page fault loop is responsible for managing UFOs as they are accessed.
This involves loading and unloading UFOs fragments in and out of memory,
in response to the needs of the user application.  It provides mechanisms for
populating areas of memory, a garbage collector for UFO fragments, and a system
for persistently caching modified fragments.  The user does not interact with
the page fault loop directly. Instead, the page fault loop is registered as a
handler for page faults with the Linux kernel for a range of virtual memory
addresses. The subsystems of the page fault loop are always reactions to
operations performed on memory guarded by the UFO core framework.

\subsection{Objects}

These userfault objects are user-facing, logical structures representing
complete larger-than-memory objects of a host language.  Logically, each
UFO owns a range of consecutive addresses whose contents are defined by a
single, specific, user-defined \emph{population function}.

\begin{figure}[t]
  \centering
  \includegraphics[width=\linewidth]{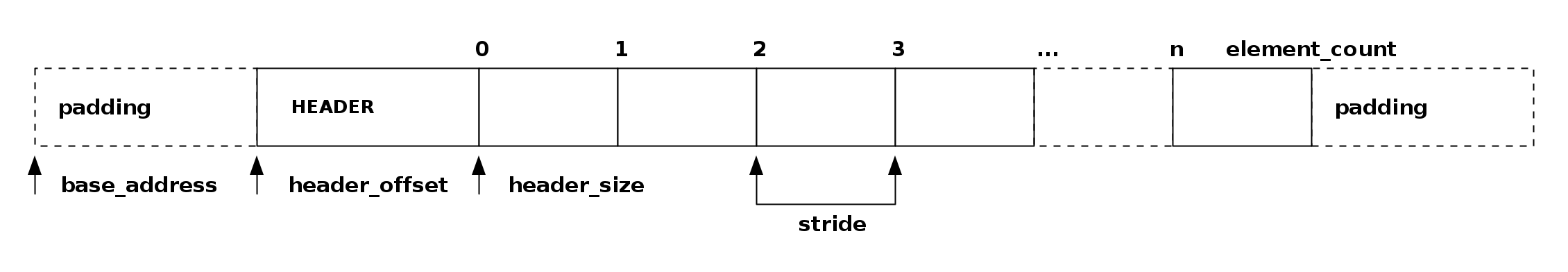}
  \caption{UFO layout.}
  \label{fig:ufo-layout} 
\end{figure}

While UFO core is agnostic with respect to the layout of host language
objects, we apply a simplifying assumption toward their internal representation
to facilitate the definition of population functions for fragments
of objects. We assume that UFOs represent arrays, each containing a header
followed by a body consisting of some number of indexed, uniformly-sized elements.
We show the logical layout of a UFO in Fig.~\ref{fig:ufo-layout}.
The boundary between the header and the elements is immutable and falls at the
boundary between the first and second segment of the UFO. The front of the UFO
is padded to accommodate the boundary. The rear is padded to align the UFO with
page size.
The header is initially empty and its contents are not generated by UFO core.
The contents of elements in the body are generated by the population
function.

The population function is executed during the lifecycle of the UFO to provide
contents of elements as they are accessed. The definitions of population
functions are external to UFO core. 
The function generates the contents for a range of elements for a specific UFO,
where the first and last index of the generated elements are specified via function
parameters. 
UFO core may demand that the function populate any contiguous region within a
UFO. A generated region may overlap other regions, and regions may be populated in any
order as well as re-populated repeatedly. To accomodate this behavior, populate
functions must be deterministic and idempotent.
Since the state of the host runtime is unknown at the time of any specific
memory access, population function must be careful about interacting with the
host runtime.
Currently population function may not attempt to access other UFOs, since it
would lead to nested userfault events.
We show an example population function in
Fig.~\ref{fig:population-function-example}.

\subsection{Segments}

\begin{figure}[t]
{ \footnotesize
  \centering
  \begin{verbatim}
  typedef struct { int from; int to; int by; } ufo_seq_data_t;
  int populate_sequence(uint64_t start_ix, uint64_t end_ix, 
                        ufUserData ufo_ud, char* target) {
    ufo_seq_data_t* data = (ufo_seq_data_t*) ufo_ud;
    for (size_t i = 0; i < end_ix - start_ix; i++) {
      ((int *) target)[i] = 
          data->from + data->by * (i + start_ix);
    }
    return 0;
  }
  \end{verbatim}
  \caption{Population function: from-to-by sequence.}
  \label{fig:population-function-example}
}    
\end{figure}

Internally, UFOs are split into \emph{segments}, each segment representing a
manageable chunk of the object's address range. At any point any segment can be
actively held in memory (\emph{materialized}) or be removed from memory
(\emph{dematerialized}). Dematerializing segments either destroys or caches
data, depending on circumstances. Materializing a segment involves (re)generating
its data through its population function or retrieving the data from a
pre-existing cache. 
Segment management is entirely transparent to the end user.

UFO core has no way of tracking accesses to segments after they are
materialized.  Therefore, ensuring that written values are not forgotten at
dematerialization requires caching.  Dirty segments are detected by comparing
the hash of their contents at the time of dematerialization with the hash after
their most recent materialization. Hashes are computed using the 256-bit BLAKE3
algorithm \cite{CANH20}. Dematerialization of dirty segments will first cause
their contents to be stored in anonymous temporary persistent storage. Each UFO
has its own file which remains in existence as long as the UFO is alive. All
cache files are cleaned up at program termination by the operating system.


UFO core keeps count of how much memory is being used by materialized segments.
UFO core carries two user-defined parameters: the high and low
water marks. When the amount of memory taken up by materialized segments
exceeds the high water mark, the UFO garbage collector is called and it starts
dematerializing segments until the low water mark is reached.

The garbage collector walks the loaded segments queue (implemented as a
circular buffer) starting from the longest residing segment. It dematerializes
segments one by one until enough space has been freed. Dematerialization does
not immediately destroy the area of memory. Instead, the kernel is signaled
that the page is no longer in use and can be recycled. The kernel lowers the
resident set size immediately and but recycles the memory at its convenience.

\section{R UFO API}

\begin{figure}[t]
  \centering
  \includegraphics[width=\linewidth]{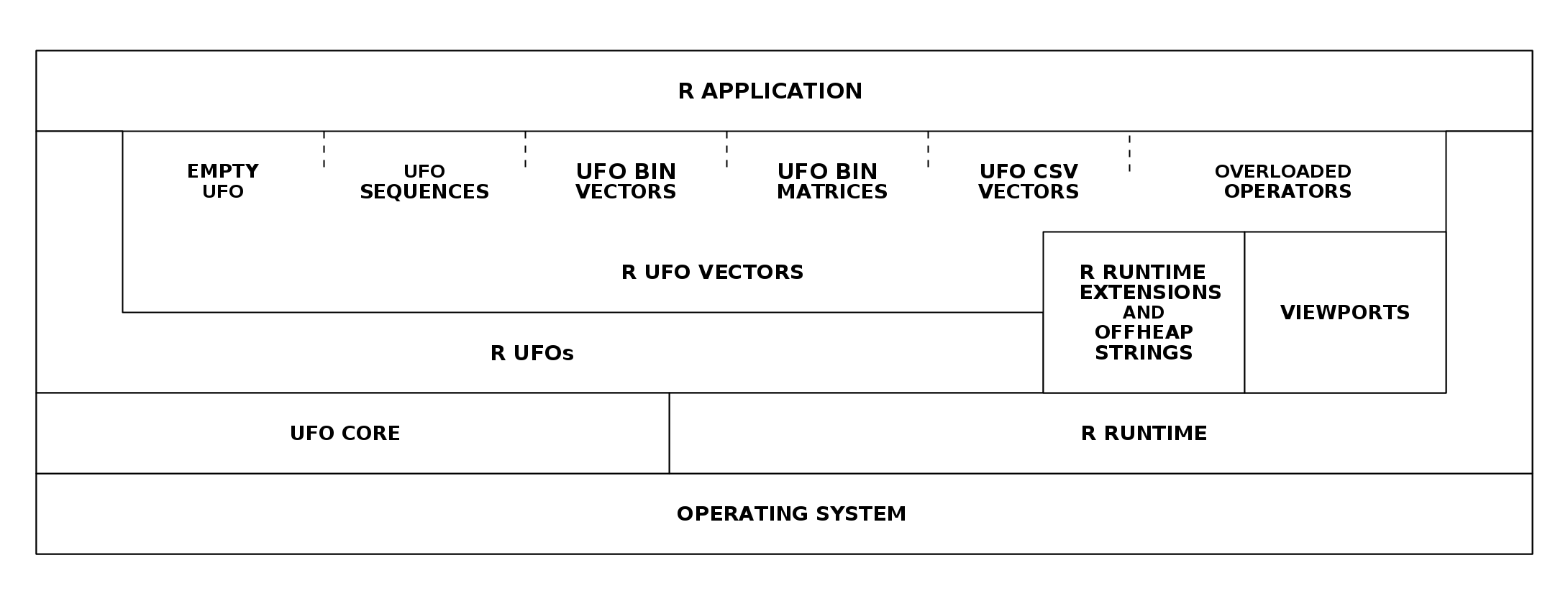}
	\caption{R UFO API architecture diagram.}
  \label{fig:sandwich} 
\end{figure}


We implemented language-specific UFO API for the R language. We picked the R
language because it is used by data scientists in fields like computational
biology, statistics, artificial intelligence, and machine learning, which deal
with large volumes of data, often represented as either vectors or data frames
(tables with uniformly sized vectors representing columns---à la CSV files).
The R ecosystem contains a many larger-than-memory libraries that create
object-oriented abstractions to hide the details of memory management from end
users while transparently representing the data to the programmer as vectors or data frames
\cite{KEW13,df,PAZB18,BV17,dbplyr}.

The R UFO API has two levels of services (Fig.~\ref{fig:sandwich}). The R UFOs
library ties the UFO core framework into the R runtime, providing an API to
specific R vector back-ends. It provides a constructor that creates a UFO
with a specific user-defined population function. It does so by plugging into
R's custom allocator API (and garbage collector),
replacing calls to \verb|malloc| and \verb|free| with calls to the R UFO core
event API. The UFO allocator returns an area of
memory to the R runtime, which is the runtime populates with an appropriate header.
In most cases, the R runtime will not pre-fill the vector, but if this happens,
UFOs ignore specific writes and their population function generates the
appropriate pre-fill values.

We provide four back-end implementations built on top of R UFOs.
Binary file-backed vectors (and matrices) read data directly from a binary
file. The data is located via seeking. CSV vectors each read a single column of
a CSV file. The values are parsed on-the-fly from a fragment of a pre-scanned
CSV file.  From-to-by sequences lazily generate data from a simple formula,
based on the index of an element. Empty vectors are pre-filled with a default
value and can be used to store large intermediate results of
computation. 

The biggest difficulty in implementing R vectors, is that R operations do not
allow custom allocation to be used in the results of arithmetic operations and
many functions. For this reason, in addition to back-end implementations, R UFO
API also provides a reimplementation of R operators that write results to UFOs,
as well as a toolkit for chunking the execution of existing functions while
aggregating the results into a UFO.

\section{Performance}

\begin{figure*}
    \begin{subfigure}{.33\textwidth}
        \includegraphics[width=.89\textwidth]{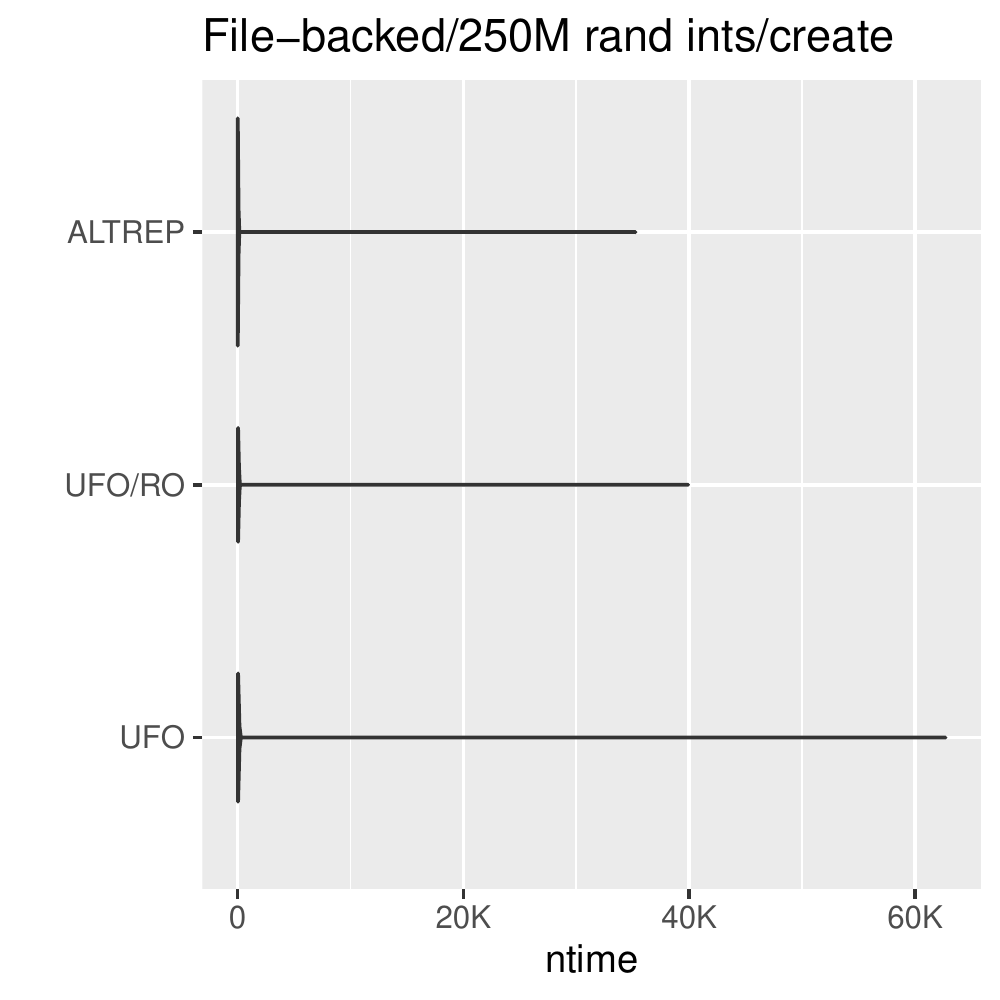}
    \end{subfigure}
    \begin{subfigure}{.33\textwidth}
        \includegraphics[width=.89\textwidth]{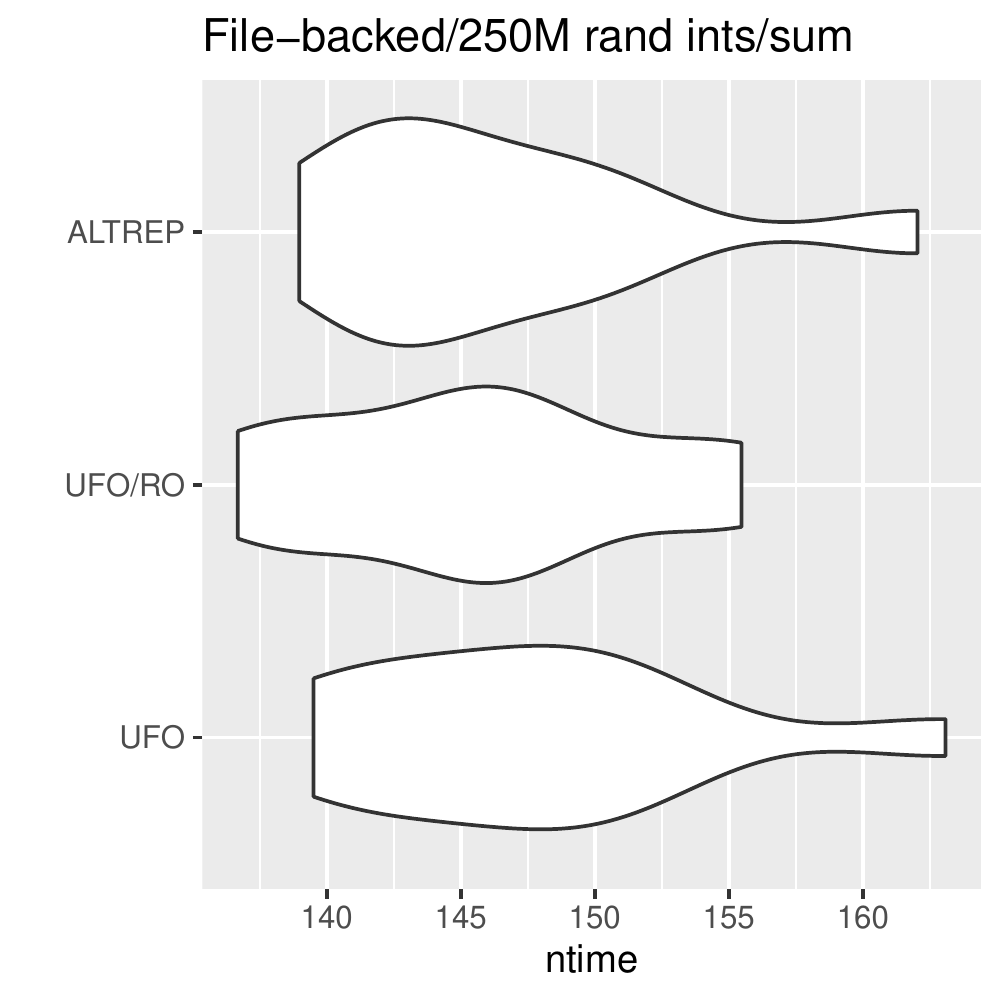}
    \end{subfigure}
    \begin{subfigure}{.33\textwidth}
        \includegraphics[width=.89\textwidth]{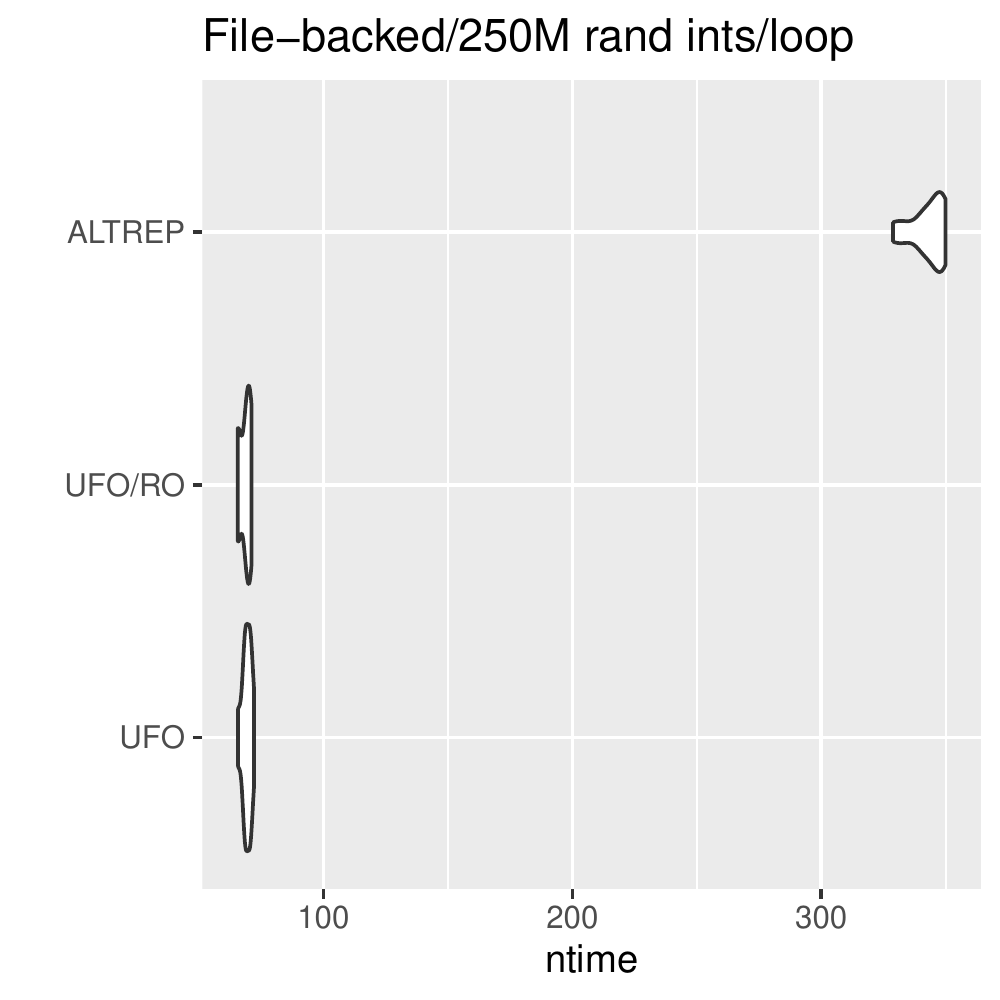}
    \end{subfigure}

    \begin{subfigure}{.33\textwidth}
        \includegraphics[width=.89\textwidth]{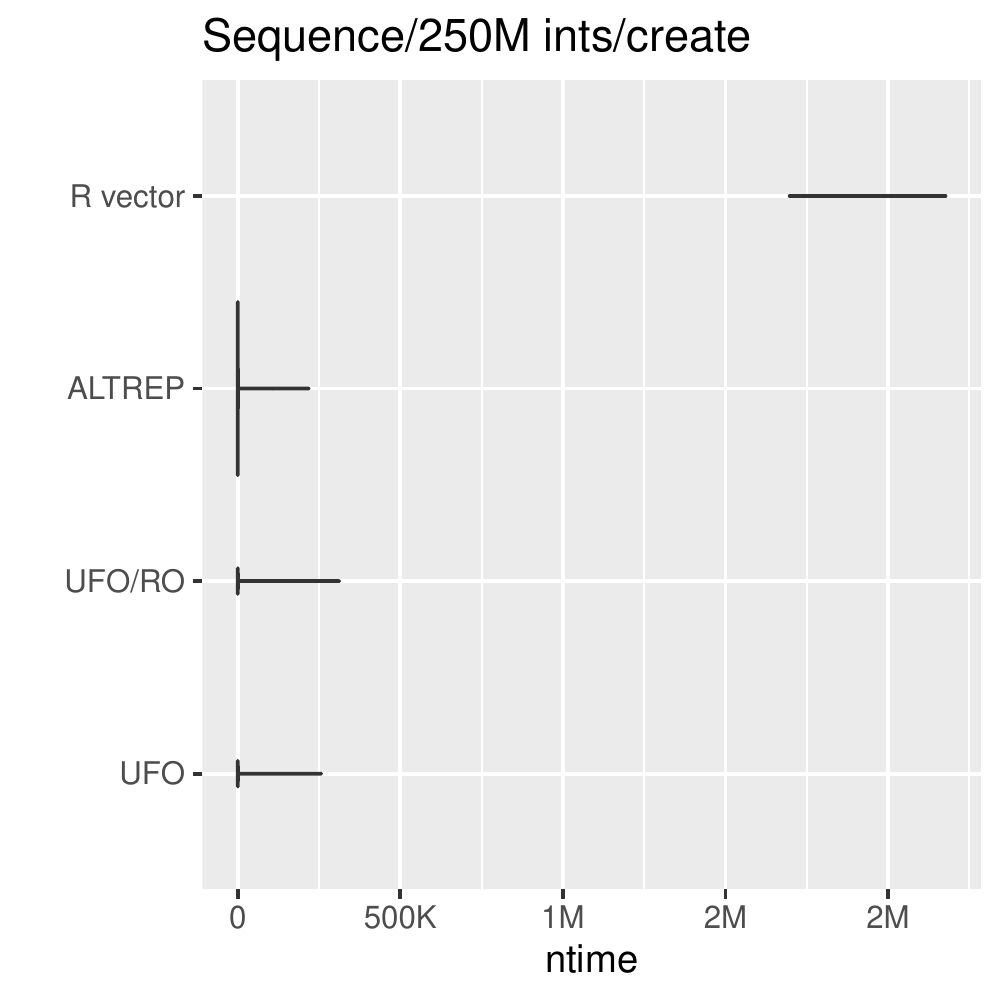}
    \end{subfigure}
    \begin{subfigure}{.33\textwidth}
        \includegraphics[width=.89\textwidth]{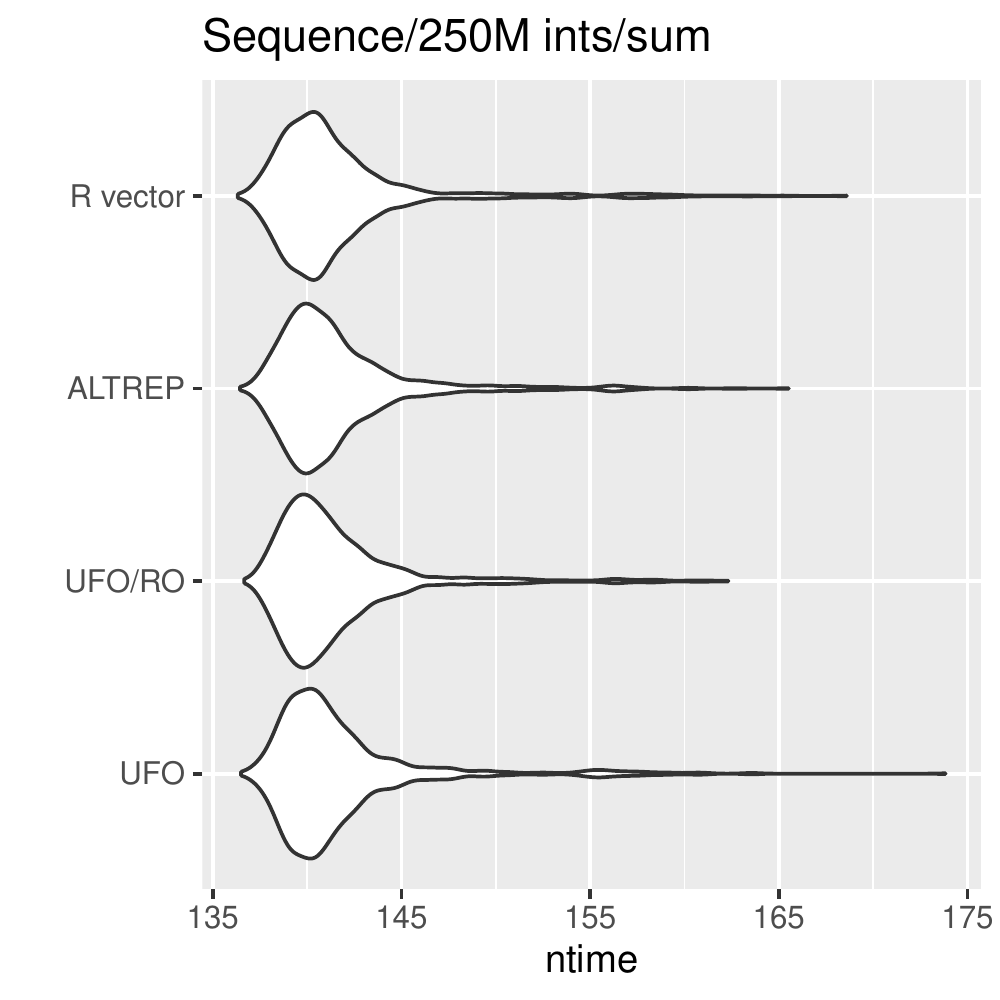}
    \end{subfigure}
    \begin{subfigure}{.33\textwidth}
        \includegraphics[width=.89\textwidth]{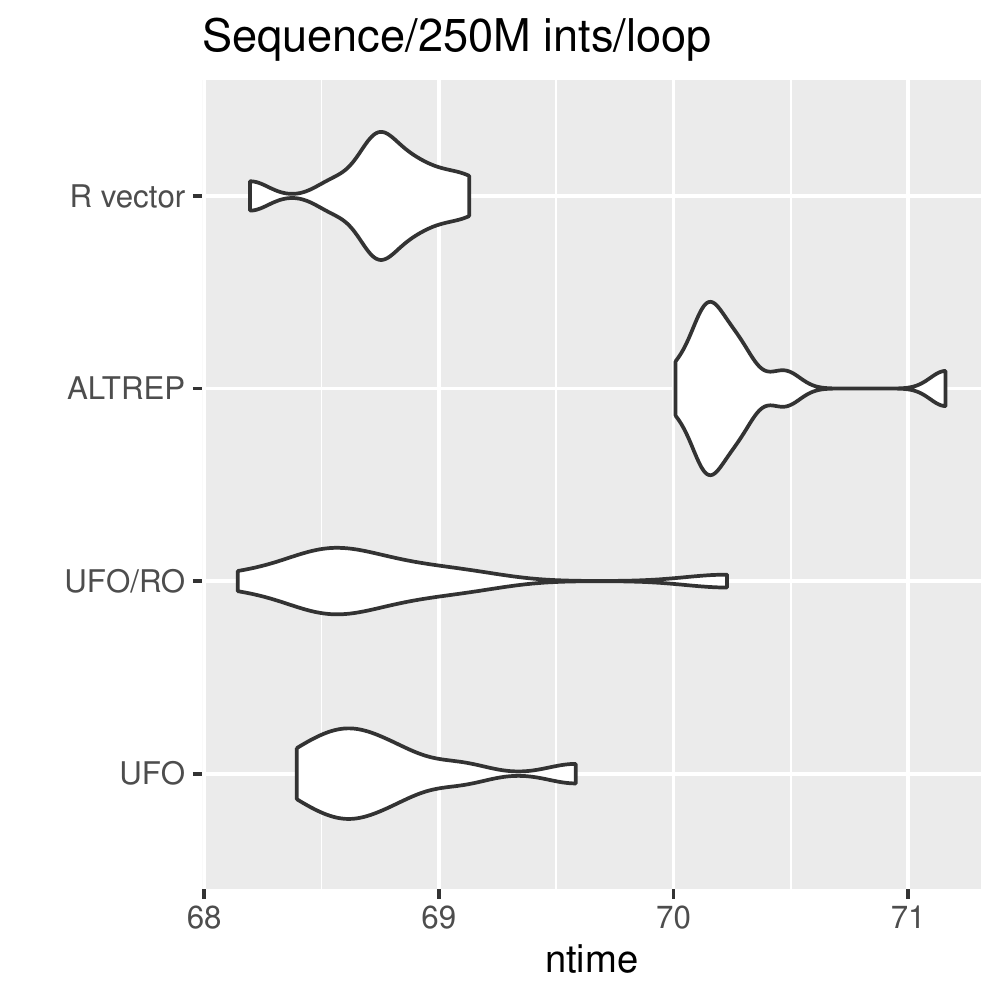}
    \end{subfigure}
	\caption{Performance evaluation.}
  \label{fig:performance}
\end{figure*}


We benchmark UFO performance measured against ALTREP and standard R vectors. 
ALTREP is a good candidate for comparison because it represents frameworks that
create an object-oriented--like facade over complex functionality while
appearing as simple vectors. ALTREP is integrated into the R
runtime, giving it a performance edge over user-created libraries.
%
We test UFOs in two modes: read/write mode and read-only
mode. Read-only mode does not persist changes done to UFOs, which removes the
need to calculate hashes of segment contents when loading and unloading them.

We use two identically implemented back-ends for UFOs and ALTREP.
File-backed vectors read 4-byte integers from a binary file on disk by seeking
to the position of the vector and reading one or more consecutive values.
This back-end has a relatively high overhead of retrieving a single
value, which can be amortized by populating entire regions at once.
Sequence vectors represent from-to-by sequences calculated on the fly (see
Fig.~\ref{fig:population-function-example}).
Computing an individual element of the sequence is cheap.
We measured the time
it takes to create a 1GB vector (1K iterations), calculate the sum of its
contents (1K iterations), and execute an identity function on each of its
elements (10 iterations).

We ran the experiments on a machine with an Intel Core™ i7-10750H CPU @
2.60GHz$\times$12 process, 32GB RAM and a and a Samsung SSD 970 EVO 500GB drive
running 64-bit Ubuntu 20.10 with 5.8.0-53-generic Linux kernel. 
We show the results of the evaluation in Fig.~\ref{fig:performance}. Each plot
shows the results for either the creation, sum, or loop microbenchmark. The top
row shows results for the file-backed vectors, the bottom one for sequences.
The X-axis always shows vector implementations and Y-axis show execution time
in nanoseconds. The results are plotted as a violin plots showing the
distribution of execution times over multiple iterations.

We observe that UFOs and ALTREP have similar performance for vector creation
and the execution time is negligibly small for both frameworks, with some
outliers we attribute to initialization and garbage collection. The startup
time is higher for R vectors implementing a sequence, because the vector must
populated up front, as opposed to UFOs and ALTREP, which calculate these values
on demand. This initialization cost for standard vectors could eventually be
amortized over multiple passes over the vector.

Sums also yield similar performance for all frameworks. The lightweight
calculation overhead involved in sequences especially washes away performance
differences. 
For file-backed vectors UFOs and ALTREP also perform similarly.  The R runtime
calculates the sum of a vector using a fast arithmetic function. This function
cooperates with ALTREP to chunk the vector into regions, which allows ALTREP
to amortize the overhead of preparing a file for reading and seeking.  While
the R runtime does not similarly chunk the execution for UFOs, the UFO
framework makes sure to read no less than 1MB of elements at-a-time and cache
data, yielding a similar amortization. Thus, the performance for both
frameworks is similar.  When the hashing mechanism is turned off for read-only
UFO vectors, a significant overhead cost is removed for UFOs, yielding a small,
but visible improvement in performance.

An importance difference in performance between UFOs and ALTREP stems from the
fact that ALTREP performs dynamic dispatch whenever values are accessed, be it
a region or a single value. The R runtime attempts to turn individual value
accesses into region accesses for ALTREP, but this can only works for specific
operations. When the loop benchmark executes, it always executes a function on
a single value from a vector, leading to repeated dispatch in ALTREP, and so,
deteriorates performance significantly. UFOs also have set-up costs relating to
loading data for an accessed value, however these costs are always amortized by
loading an entire segment into memory. This gives UFOs an advantage over
ALTREP's dispatch and produces performance close to ordinary vectors when
consecutive elements are accessed. However, this approach is costly if
the access pattern is spread out, causing the UFO to load and
unload a segment for each single value read.


\section{Conclusions and future work}

The UFO framework explores avenues of cooperating with the operating system to
use memory in non-traditional ways. We implement a framework that uses user
faults to lazily provide data to a language's runtime object. This allows the
implementation of structures that generate data from a variety of sources, but
follow the memory layout of standard runtime objects, so they can be
introspected safely.  Nevertheless, they can implement complex back-ends and
provide access to larger-than-memory data that never needs to materialize into
memory fully. Implementing objects via userfaults also has an impact on
performance as overhead is amortized over loading large segments of data and
the host runtime can rely on direct memory accesses into userfault object.

Future work includes implementing a mechanism for supporting recursive calls
between UFOs and reacting to specific memory access patterns to limit
unnecessary memory usage. We would also like to explore the applicability of
this approach outside of the Linux ecosystem and in other language runtimes.




\begin{acks}
This work is supported by the Czech Ministry of
Education, \linebreak 
Youth and Sports from the Czech Operational Programme Research,
Development, and Education, under grant agreement No.
\linebreak CZ.02.1.01/0.0/0.0/15\_003/0000421 and the European Research Council (ERC)
under the European Union's Horizon 2020 research and innovation programme (grant
agreement No. 695412).
\end{acks}

\bibliographystyle{ACM-Reference-Format}
\bibliography{bibliography}

\end{document}